\documentclass[12pt,preprint]{aastex}
\usepackage{emulateapj5,pstricks}

\newcommand{\be}{\begin{equation}}
\newcommand{\ee}{\end{equation}}
\newcommand{\ba}{\begin{eqnarray}}
\newcommand{\ea}{\end{eqnarray}}

\shorttitle{}
\shortauthors{Wang \& Freese}

\begin{document}

\title{Probing Dark Energy Using Its Density \\
Instead of Its Equation of State}

\author{Yun~Wang$^{1}$, and Katherine Freese$^{2}$}
\altaffiltext{1}{Department of Physics \& Astronomy, Univ. of
  Oklahoma, 440 W Brooks St., Norman, OK 73019; email:
  wang@nhn.ou.edu} \altaffiltext{2}{Michigan Center for Theoretical
  Physics, Univ. of Michigan, Ann Arbor, MI 48109; email:
  ktfreese@umich.edu}
\begin{abstract}
  
  The variation of dark energy density with redshift, $\rho_X(z)$,
  provides a critical clue to the nature of dark energy.  Since
  $\rho_X(z)$ depends on the dark energy equation of state $w_X(z)$
  through an integral, $\rho_X(z)$ can be constrained more tightly than
  $w_X(z)$ given the same observational data.  We demonstrate this
  explicitly using current type Ia supernova (SN Ia) data [the
  Tonry/Barris sample], together with the Cosmic Microwave Background
  (CMB) shift parameter from CMB data (WMAP, CBI, and ACBAR), and the
  large scale structure (LSS) growth factor from 2dF galaxy survey
  data. We assume a flat universe, and use Markov Chain Monte Carlo 
  (MCMC) technique in our analysis.  
  We find that, while $w_X(z)$ extracted from current data is
  consistent with a cosmological constant at 68\% C.L., $\rho_X(z)$
  (which has far smaller uncertainties) is not.  Our results clearly
  show the advantage of using $\rho_X(z)$, instead of $w_X(z)$, to
  probe dark energy. 

\end{abstract}

%\end{document}


\section{Introduction}

Recent observations of type Ia Supernovae \citep{Riess98,Perl99} indicate that
the universe is accelerating.  A fundamental quest in physics and
cosmology is to identify the nature of the ``dark energy'' 
driving this acceleration.
Possibilities include: (1) a cosmological constant, (2) a time dependent
vacuum energy, or a scalar field known as ``quintessence'' that
evolves dynamically with time 
\citep{fafm,peebles88,Wett88,frieman,caldwell98,Zlatev99}
\footnote{See \cite{Pad03} and 
\cite{peebles03} for reviews with more complete lists of references).} 
or (3) modified Friedmann equation, e.g. the Cardassian models 
\citep{freeselewis,freese03,mpcard,Wang03}, that
could result as a consequence of our observable universe living as a
3-dimensional brane in a higher dimensional universe.  Other proposed
modifications to the Friedmann equation include 
\cite{Parker99,ddg,Bilic02,Ahmed02,Capo03,Carroll03b,Meng03,Puet04}.  The
various dark energy models produce dark energy densities $\rho_X(z)$
with different redshift dependences.  Hence, in order to differentiate
between dark energy models, it is important that we allow the
dark energy density to be an arbitrary function of redshift $z$
\citep{Wang01a,Wang01b,Wang03}.

A powerful probe of dark energy is type Ia supernovae (SNe Ia), which
can be used as cosmological standard candles to measure how distance
depends on redshift in our universe.  The luminosity distance $d_L(z)
= (1+z) r(z)$, with the comoving distance $r(z)$ given by 
\begin{equation} 
r(z)= cH_0^{-1} \int_0^z \frac{dz'}{E(z')}, 
\end{equation} 
with 
\begin{equation}
\label{eq:E(z)}
E(z) \equiv \left[ \Omega_m(1+z)^3 + \Omega_k (1+z)^2 + \Omega_X
  \rho_X(z)/\rho_X(0) \right]^{1/2}, 
\end{equation} 
where $\Omega_k \equiv
1-\Omega_m-\Omega_X$, and $\rho_X(z)$ is the dark energy density.  

The dark energy equation of state, $w_X(z)$, is related to $\rho_X(z)$
as follows \citep{Wang01a}: 
\begin{equation}
\label{eq:wrhoprime}
w_X(z) =\frac{1}{3}(1+z)
\frac{\rho'_X(z)}{\rho_X(z)} -1,  
\end{equation} 
so that
\begin{equation}
\label{eq:rhoprimew}
\frac{\rho_X(z)}{\rho_X(0)} = \exp\left\{ \int_0^z \frac{3
    [1+w_X(z)]}{1+z} \right\}.  
\end{equation} 
One can see that it is easier to extract $\rho_X(z)$ from the data
than to extract $w_X(z)$.  To obtain the dark energy density directly,
one need only take a single derivative of the luminosity distance,
whereas to extract $w_X(z)$, one needs to take a second derivative as
well; from Eq.(\ref{eq:wrhoprime}) one can see that $w_X(z)$ is on the
same footing as $\rho_X'(z)$.  Specifically, \cite{Wang01a} argued
that $\rho_X(z)$ should be preferred since it suffers less from the
smearing effect (due to the multiple integrals that relate $w_X(z)$
to $d_L(z)$) that makes constraining $w_X(z)$ extremely difficult
\citep{Maor01,barger01}.  \cite{Tegmark02} came to the same conclusion.
However, researchers have generally chosen to parametrize dark energy
using its equation of state $w_X(z)$. Some have used
  $H(z)=H_0 E(z)$ (for example, see \cite{Kujat02,Daly03,peri}, and
  references therein), which is similar to $\rho_X(z)$, but
  measurements of which are not as straightforward to interpret, since
  $E(z)$ depends on $\Omega_m$ (see Eq.(\ref{eq:E(z)})).

In this paper, we explicitly demonstrate 
the advantage of using $\rho_X(z)$, instead 
of $w_X(z)$, to probe dark energy. 
Sec.2 contains a comparison of $w_X(z)$ and $\rho_X(z)$
parametrizations using current SN Ia, CMB, and LSS data.
We give a recipe for parametrizing dark energy using $\rho_X(z)$
in Sec.3.
Sec.4 contains a summary and discussions.

\section{Dark energy equation of state versus \\
dark energy density}
%\section{\mbox{$w_X(z)$} versus \mbox{$\rho_X(z)$}}

>From SN, CMB, and LSS data, we independently reconstruct
first the dark energy equation of state and then the dark 
energy density directly.
We use (1) current SN Ia data from Tonry et al. 2003
and Barris et al. 2003 [the Tonry/Barris sample], 
(2) the Cosmic Microwave 
Background (CMB) shift parameter \citep{Bond97}
from CMB data [WMAP \citep{Bennett03,Spergel03}, CBI \citep{Pearson03}, 
and ACBAR \citep{Kuo02}], and 
(3) the large scale structure (LSS) growth factor 
from 2dF \citep{Percival02,Verde02,Hawkins03} galaxy survey data.\citep{Wang04}

We parametrize the data first in terms of the dark energy equation of state,
and then in terms of the dark energy density, to see which
parametrization produces the reconstruction with the least
uncertainty. 

To find the function $w_X(z)$ which best fits the data, we consider a
five-dimensional parameter space: the function evaluated at three
discrete redshift intervals, the value of $\Omega_m$, and the value of
the Hubble constant.  We assume a flat universe.  It is our goal to
find a set of values for these five parameters that fits the
data.  We use the Markov Chain Monte Carlo (MCMC) technique
\citep{neal,LB02}, which selects randomly from the 5D parameter space,
evaluates $\chi^2$, and create a large number of sets of parameter
values (each set is a MCMC sample); 
we use $10^6$ MCMC samples \footnote{   
Here $\chi^2$ is only used to move around efficiently in the 
entire parameter space (based on entropy considerations), 
such that for sufficent sampling, the resultant parameter
distributions converge to the true probability distribution functions (pdf's).
This leads to smooth pdf's since they receive contributions
from all MCMC samples. }.  For the current data, we consider
the function $w_X(z)$ at three redshift values: $z$=0, $z_{max}/2$,
and $z_{max}$ (where $z_{max}$ is the maxmimum redshift of SNe
Ia)\footnote{This is the largest number of values one can get out of
  the current (sparse) data.}. We find the values (that fit the
data) at these points and interpolate at all intermediate redshifts.
The parameters estimated from data are $w_X(0)$, $w_X(z_{max}/2)$,
$w_X(z_{max})$, $\Omega_m$, and a dimensionless Hubble constant $h$.

Fig.1a shows the $w_X(z)$ reconstructed from
192 SNe Ia from the Tonry/Barris sample, 
\footnote{Flux averaging has been performed to reduce the bias
in estimated parameters due to weak gravitational lensing.
See \cite{Wang00b} and \cite{Wang04} for details.}
combined with CMB (shift parameter 
${\cal R}_0 = 1.716 \pm 0.062$)
and LSS data (growth parameter $f_0 \equiv 
f(z=0.15) =0.51\pm 0.11$)\citep{Wang04}.
The regions inside the solid and dashed lines correspond
to 68.3\% and 95\% confidence levels 
respectively;
the 68.3\% confidence level (C.L.) region is also shaded.
The circles indicate the mean values 
at the three redshift points, $z$=0, $z_{max}/2$, and $z_{max}$.
The other simultaneously estimated parameters 
(mean, 68.3\% and 95\% confidence ranges) are:
$\Omega_m=.39 [.29, .50] [.21, .57]$ and
$h= .658 [.642, .674] [.627, .689]$.\footnote{The uncertainty 
on $h$ are statistical error only, not including the contribution
from the much larger SN Ia absolute magnitude error of 
$\sigma_h^{int}\simeq 0.05$.\citep{Wang04}} 
Clearly, the equation of state is consistent with a constant
$w_X(z)=-1$ for all redshifts at 95\% confidence level (C.L.).  At
68.3\% C.L., it is consistent with a constant for $0 \la z \la 0.5$
and marginally consistent with $w_X(z)=-1$ for $0.5 \la z \la 1$.

Fig.1b shows the $\rho_X(z)$ directly reconstructed from the same data
as Fig. 1.  The same technique of discretizing the function $\rho_X(z)$
has been used.  The solid lines and dashed lines indicate the 68.3\% and
95\% confidence levels respectively 
\citep{Wang04}. The other simultaneously estimated
parameters (mean, 68.3\% and 95\% confidence ranges) are:
$\Omega_m=.33 [.27, .39] [.22, .46]$, $h= .660 [.644, .673] [.630,
.688]$.  One can see that the uncertainties in Fig.1b on $\rho_X(z)$
obtained from the data
are smaller than those on $w_X(z)$ obtained from the data.
We see that the time dependence of the dark energy density
deviates from a constant at 68.3\% C.L. (a similar statement
could not be made from the $w_X(z)$ reconstruction).
With more data in the future, the statistical significance of this
discrepancy will become more clear.

For comparison, in Fig.2 we have also plotted $\rho_X(z)$ obtained
in a more indirect way: by first obtaining $w_X(z)$ from the data, as
described above, and then integrating over redshift as in
Eq.(\ref{eq:rhoprimew}).  Clearly the uncertainties obtained in this
way are far larger than if one obtains the dark energy directly from
the data.  While the results from $w_X(z)$ parametrization and
$\rho_X(z)$ parametrization \citep{Wang04} are consistent with one
another, the $w_X(z)$ parametrization results have uncertainties that
are several times larger.  One can also obtain $w_X(z)$ indirectly
from $\rho_X(z)$ in Fig.1b (similar to what was done
by \cite{alam}). However, doing so would require taking
the derivative of the polynomial used in the interpolation,
thus making the resultant $w_X(z)$ dependent on the interpolation
technique used.

Our main result is that one can learn more information by
reconstructing $\rho_X(z)$ rather than $w_X(z)$ from the data.  At
95\% C.L., both the $w_X(z)$ and $\rho_X(z)$ reconstructions are
consistent with a cosmological constant.  However at 68.3\% C.L., the
$\rho_X(z)$ reconstruction has smaller uncertainties
and hence shows more information than the $w_X(z)$
reconstruction: the $\rho_X(z)$ reconstruction is {\it not}
consistent with a time-independent dark energy.  Even with the
$\rho_X(z)$ parametrization, a significant number of SNe Ia at $z>1$
from a deep SN survey on a dedicated telescope \citep{Wang00a} will be
required to place robust constraints on the time-dependence of
$\rho_X(z)$.

\section{A recipe for parametrizing dark energy \\
using its density}
%\mbox{$\rho_X(z)$}}

For the convenient application of our methodology by others,
we now present a recipe for parametrizing dark energy using $\rho_X(z)$
as an arbitrary continuous function.

(1) Choose the number of redshift bins, $N$ (the number of parameters  
for $\rho_X(z)$).
$N$ needs to be sufficiently large to probe the time-variation
of $\rho_X(z)$. However, if $N$ is too large, the uncertainties
on all the estimated parameters will increase, leading
to less stringent constraints. $N=2$ is appropriate
for current (sparse) data.

(2) The values of the dimensionless dark energy density
$f_i \equiv \rho(z_i)/\rho_X(0)$ ($i=1,2,...,N)$ are the independent variables
to be estimated from data. 
Note that $z_N=z_{max}$ (the maximum redshift of SNe Ia in the data).

(3) Parametrize $\rho_X(z)/\rho_X(0)$ as a continuous function,
given by interpolating its amplitudes at equally spaced
$z$ values in the redshift range covered by SN Ia data
($0 \leq z \leq z_{max}$),
and a constant at larger $z$ ($z>z_{max}$,
where $\rho_X(z)$ is only weakly constrained by CMB data).
The results should not be sensitive to the interpolation
method used. 
Polynomial interpolation was
used in \cite{Wang04}. For $N=2$, this gives
\be
\label{eq:rho_X}
\frac{\rho_X(z)}{\rho_X(0)}=1+ \left(4 f_1 -f_2 -3\right)\, 
\frac{z}{z_{max}} + \left(f_2 -2 f_1 +1\right)\, 
\frac{2z^2}{z^2_{max}},
\ee
where $f_1=\rho(z_{max}/2)/\rho_X(0)$, and
$f_2=\rho(z_{max})/\rho_X(0)$.

(4) Use Eq.(\ref{eq:rho_X}) or its equivalent (if the
interpolation method or $N$ differs) in
all equations where the factor $E(z)$ from
Eq.(\ref{eq:E(z)}) appears.

{\it Caution:} It is important to note that we are
using a polynomial to interpolate $\rho_X(z)$
between equally spaced $z$ values; the independent
variables are the values of $\rho_X(z)$
at these $z$ values, as in Eq.(\ref{eq:rho_X}).
In this case, the errors on the reconstructed $\rho_X(z)$ are 
tied to how the quality of the data varies with $z$ (sparse
data lead to large errors).
Changing the interpolation method from polynomial interpolation
to a different method should have negligible effect
on the reconstructed $\rho_X(z)$. 
On the other hand, if a polynomial is used as a global
fit function with its coefficients being the independent variables;
the errors on the reconstructed $\rho_X(z)$
will not correlate with how the quality of data 
varies with $z$.

\section{Summary and Discussion}

The critical first step in solving the mystery of dark energy 
is to determine whether the dark energy density $\rho_X(z)$
varies with time.\citep{Wang01a}
A definitive answer to this question can have profound
implications for particle physics and cosmology.

Our main result is that one can learn more information by
reconstructing $\rho_X(z)$ rather than $w_X(z)$ from the data.  The
two quantities are related by an integral, which in the case of
$w_X(z)$ smears out much of the information one could otherwise learn.
We show this explicitly by using a combination of SN Ia data from the
Tonry/Barris sample as well as CMB (WMAP, CBI, and ACBAR)
and large scale structure (2dF)
data.  At 95\% CL, both the $w_X(z)$ and $\rho_X(z)$ reconstructions
are consistent with a cosmological constant.  However at 68\% CL, the
$\rho_X(z)$ reconstruction has smaller uncertainties and hence shows
information that the $w_X(z)$ reconstruction cannot: the $\rho_X(z)$
reconstruction is {\it not} consistent with a time-independent dark
energy, and the dark energy density appears to be increasing with
redshift.  Future data will be required to resolve this question.

We have shown definitively the advantage of the $\rho_X(z)$
parametrization over the $w_X(z)$ parametrization in determining the
time-variation of $\rho_X(z)$.  To help others apply the $\rho_X(z)$
parametrization, we have given a recipe for using the $\rho_X(z)$
parametrization in data analysis to probe dark energy (see Sec.3).
Our methodology should be very useful in all data analysis aiming at
unraveling the nature of dark energy.

\acknowledgements 
We are grateful to Pia Mukherjee for helping us set up MCMC
and for helpful discussions,
and thank P. Gondolo for useful discussions.
This work is supported in part by NSF CAREER grant
AST-0094335 (YW) and by the DOE and the MCTP via grants at the
University of Michigan (KF).

%\clearpage
\setcounter{figure}{0} \plotone{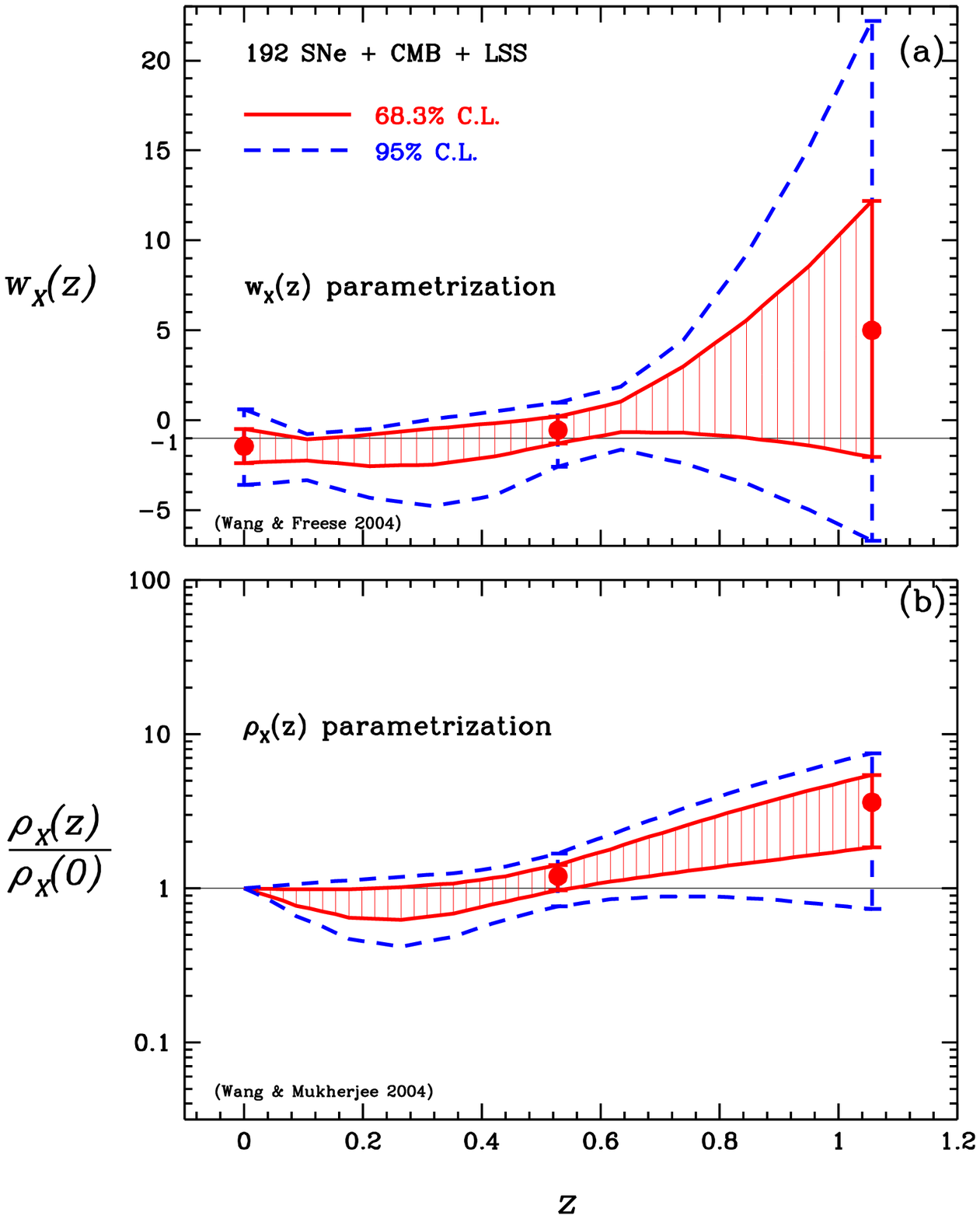} \figcaption[f1.eps] {(a) The
  $w_X(z)$ reconstructed from 192 SNe Ia from the Tonry/Barris sample,
  flux-averaged, and combined with CMB and LSS data.  The regions
  inside the solid and dashed lines correspond to 68.3\% and 95\%
  confidence levels respectively; the 68.3\% confidence level region
  is also shaded. Circles indicate the mean values of the regions.
  (b) The $\rho_X(z)$ reconstructed from the same data as Fig.1(a),
  with the same shading and line types \citep{Wang04}.  Whereas the
  $w_X(z)$ reconstruction is consistent at 68.3\% C.L. with a
  cosmological constant, the $\rho_X(z)$ reconstruction is not.  }

\setcounter{figure}{1} \plotone{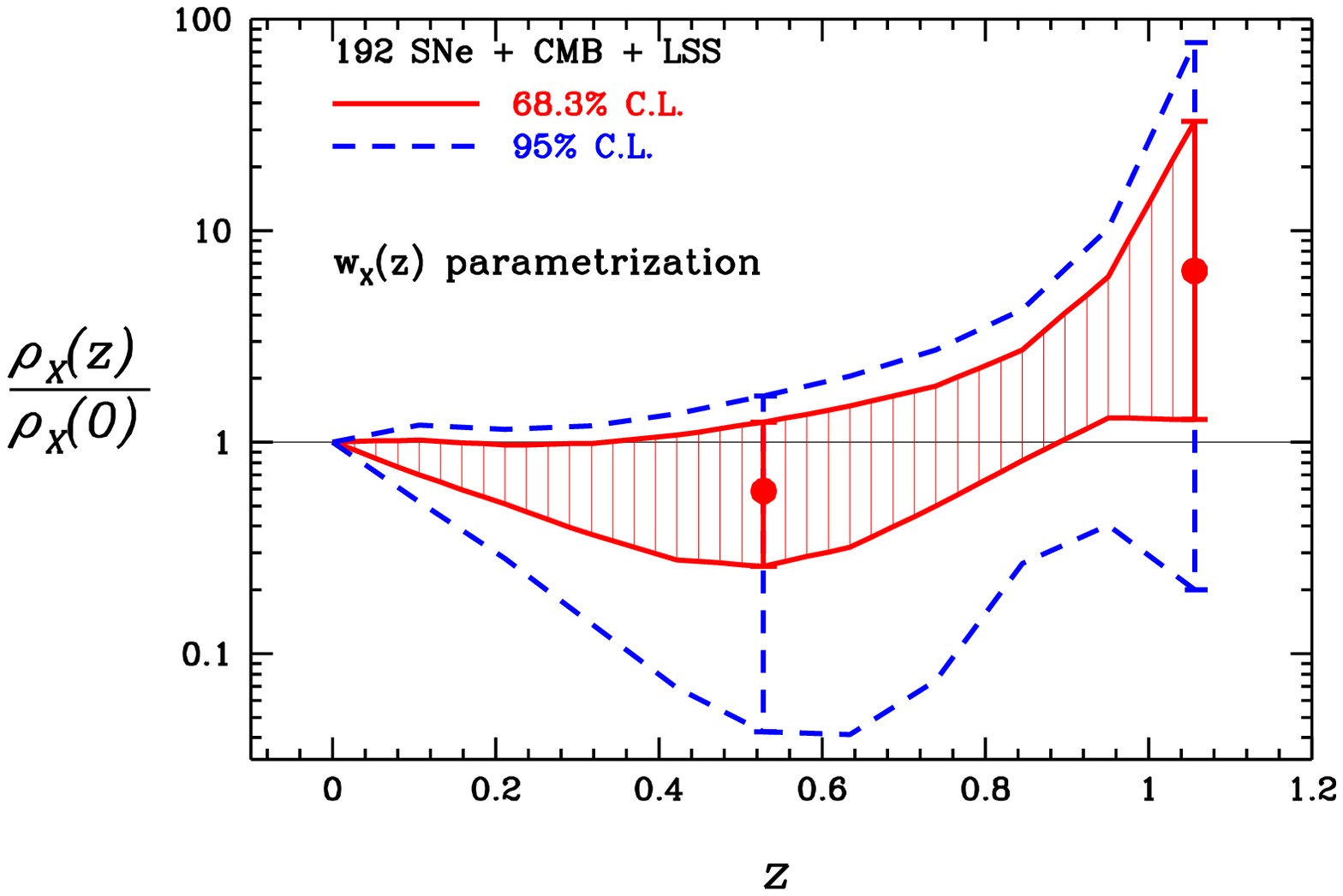} \figcaption[f2.eps] { The
  $\rho_X(z)$ reconstructed by taking the integral in Eq.(4) of the
  $w_X(z)$ plotted in Fig.1(a).  Again, we use the same data, with
  the same shading and line types.  Clearly the uncertainties in this
  method are much greater than if one obtains $\rho_X(z)$ directly
  from the data as in Fig.1(b).  }


\begin{thebibliography}{}

\bibitem[Alam et al.(2003)]{alam}
Alam, U., Sahni, V., Saini, T.D., and Starobinsky, A.A. 2003,
astro-ph/0311364

\bibitem[Ahmed et al.(2002)]{Ahmed02}
Ahmed, M., Dodelson, S., Greene, P.B., and Sorkin, R. 2002, astro-ph/0209227; 

\bibitem[Armendariz-Picon, Mukhanov, \& Steinhardt(2000)]{Armenda00}
Armendariz-Picon, C., Mukhanov, V., Steinhardt, P.J. 2000,
Phys. Rev. Lett. 85, 4438

\bibitem[Barger \& Marfatia(2001)]{barger01}
Barger, V., and Marfatia, D. 2001, Phys.\ Lett.\ B498, 67-73 

\bibitem[Barris et al.(2003)]{Barris03}
Barris, B.J., et al. 2003, astro-ph/0310843, ApJ, in press


\bibitem[Bennett et al.(2003)]{Bennett03}
Bennett, C., et al. 2003, ApJ, Suppl. 148, 1

\bibitem[Bilic, Tupper, \& Viollier(2002)]{Bilic02}
Bilic, N., Tupper, G.B, and Viollier, R., Phys.Lett. {\bf B535} 17 (2002);

\bibitem[Bond, Efstathiou, \& Tegmark(1997)]{Bond97}
Bond, J.R.; Efstathiou, G.; \& Tegmark, M. 1997, MNRAS, 291, L33

\bibitem[Caldwell, Dave, \& Steinhardt(1998)]{caldwell98} Caldwell, R., Dave, R.,
Steinhardt, P.J. 1998, Phys. Rev. Lett., 80,  1582 

\bibitem[Capozziello, Carloni, \& Troisi]{Capo03}
Capozziello, S., Carloni, S.,  and Troisi, A., astro-ph/0303041;

\bibitem[Carroll, Hoffman, \& Trodden(2003)]{Carroll03}
Carroll, S.M.; Hoffman, M., \& Trodden, M. 2003,
Phys. Rev. D68, 023509

\bibitem[Carroll et al.(2003)]{Carroll03b}
Carroll,  S., Duvvuri, V., Trodden, M., and Turner, M., astro-ph/0306438.

\bibitem[Daly \& Djorgovski(2003)]{Daly03}
Daly, R.A., \& Djorgovski, S.G. 2003, ApJ, 597, 9

\bibitem[Deffayet(2001)]{ddg} Deffayet, C. 2001, Phys. Lett. B502, 199

\bibitem[Freese et al.(1987)]{fafm} Freese, K., Adams, F.C., Frieman, J.A.,
 and Mottola, E. 1987, Nucl. Phys. B287, 797

  
\bibitem[Freese \& Lewis(2002)]{freeselewis} Freese, K., and Lewis, M.,
2002, Phys. Lett., B540, 1

\bibitem[Freese(2003)]{freese03}
Freese, K., Nuclear Physics B (Proc. Suppl.) {\bf 124}, 50 (2003)
  
\bibitem[Frieman et al.(1995)]{frieman} Frieman, J.,  Hill, J., Stebbins, A.,
and Waga, I. 1995, Phys. Rev. Lett., 75, 2077

\bibitem[Gondolo \& Freese(2003)]{mpcard} Gondolo, P.,  and Freese, K. 2003,
Phys.Rev. {\bf D68} 063509
 
\bibitem[Hawkins et al.(2003)]{Hawkins03}
Hawkins, E. et al. 2003, astro-ph/0212375, MNRAS in press


\bibitem[Knop et al.(2003)]{Knop03}
Knop, R. A., et al. 2003, astro-ph/0309368, ApJ, in press

\bibitem[Kujat et al.(2002)]{Kujat02}
Kujat, J.; Linn, A.M.; Scherrer, R.J.; \& Weinberg, D.H. 2002,
ApJ, 572, 1

\bibitem[Kuo et al.(2002)]{Kuo02}
Kuo, C.L., et al. 2002, submitted to ApJ, astro-ph/0212289

\bibitem[Lewis \& Bridle(2002)]{LB02}
Lewis, A., \& Bridle, S. 2002, Phys. Rev. D, 66, 103511, astro-ph/0205436

\bibitem[Maor, Brustein, \& Steinhardt(2001)]{Maor01}
Maor, I., Brustein, R., \& Steinhardt, P.J. 2001, 
Phys. Rev. Lett., 86, 6; Erratum-ibid. 87 (2001) 049901

\bibitem[Meng \& Wang(2003)]{Meng03}
Meng, X., \& Wang, P. (2003), Class.Quant.Grav. 20, 4949

\bibitem[Neal(1993)]{neal}
Neal, R.M. 1993, Technical Report CRG-TR-93-1,
ftp://ftp.cs.utoronto.ca/pub/~radford/review.ps.gz

\bibitem[Nesseris \& Perivolaropoulos(2004)]{peri} 
Nesseris, S., \& Perivolaropoulos, L. 2004, astro-ph/0401556

\bibitem[Padmanabhan(2003)]{Pad03}
Padmanabhan, T. 2003, Physics Reports 380, 235-320 

\bibitem[Parker \& Raval(1999)]{Parker99}
Parker, L., and Raval, A. 1999, Phys. Rev. D {\bf 60}, 063512
  
  
\bibitem[Peebles \& Ratra(1988)]{peebles88}
Peebles, P. J. E.; Ratra, B. 1988, ApJ, 325L, 17

  
\bibitem[Peebles \& Ratra(2003)]{peebles03}
Peebles, P. J. E.; Ratra, B. 2003, Rev.Mod.Phys. 75, 559-606

\bibitem[Pearson et al.(2003)]{Pearson03} 
Pearson, T.J., et al. 2003, ApJ, 591, 556-574

\bibitem[Percival et al.(2002)]{Percival02}
Percival, W.J. et al. 2002, MNRAS, 337, 1068

\bibitem[Perlmutter et al.(1999)]{Perl99} Perlmutter, S., et al. 1999,  
ApJ,  517,  565


\bibitem[Phillips(1993)]{Phillips93}
Phillips, M.M., ApJ, 413, L105 (1993)

\bibitem[Puetzfeld \& Chen(2004)]{Puet04}
Puetzfeld, D., and Chen, X. 2004, gr-qc/0402026


\bibitem[Riess et al.(1998)]{Riess98} Riess, A.~G., et al. 1998,  
AJ, 116, 1009 


\bibitem[Riess, Press, \& Kirshner(1995)]{Riess95}
Riess, A.G., Press, W.H., and Kirshner, R.P., ApJ, 438, L17 (1995)

\bibitem[Riess et al.(1999)]{Riess99}
Riess, A.G., et al. 1999, AJ, 118,  2675

\bibitem[Spergel et al.(2003)]{Spergel03}
Spergel, D.N., et al. 2003, astro-ph/0302209

\bibitem[Tegmark(2002)]{Tegmark02}
Tegmark, M. 2002, Phys. Rev. D66, 103507


\bibitem[Tonry et al.(2003)]{Tonry03}
Tonry, J.L., et al. 2003, ApJ, 594, 1-24


\bibitem[Verde et al.(2002)]{Verde02}
Verde, L. et al. 2002, MNRAS, 335, 432


\bibitem[Wang(2000a)]{Wang00a} 
Wang, Y. 2000a, ApJ, 531, 676 

\bibitem[Wang(2000b)]{Wang00b}
Wang, Y. 2000b, ApJ, 536, 531 

\bibitem[Wang \& Garnavich(2001)]{Wang01a}
Wang, Y., and Garnavich, P. 2001, ApJ, 552, 445 

\bibitem[Wang \& Lovelace(2001)]{Wang01b} 
Wang, Y., and Lovelace, G. 2001, ApJ, 562, L115 

\bibitem[Wang et al.(2003)]{Wang03}
Wang, Y.;  Freese, K.; Gondolo, P.; \& Lewis, M. 2003,
ApJ, 594, 25

\bibitem[Wang \& Mukherjee(2004)]{Wang04}
Wang, Y., \& Mukherjee, P. 2004, astro-ph/0312192, ApJ submitted

\bibitem[Wetterich(1988)]{Wett88}
Wetterich, C. 1988,  Nucl. Phys. {\bf B302}, 668 
 
\bibitem[Zlatev, Wang, \& Steinhardt(1999)]{Zlatev99}
Zlatev,  I., Wang, L., and Steinhardt, P.J. 1999,
  Phys. Rev. Lett.  {\bf 82}, 896 
   
\end{thebibliography}
\end{document}